\documentclass[12pt,preprint, nofootinbib, aps]{revtex4}
\usepackage{graphicx}
\usepackage{dcolumn}
\usepackage{bm}
\usepackage{epsfig}
\usepackage{amssymb,amsmath,amsfonts,amsbsy,epsfig}
\usepackage{color}

\def\lsim{\mathrel{\rlap{\lower4pt\hbox{\hskip1pt$\sim$}}
    \raise1pt\hbox{$<$}}}         
\def\gsim{\mathrel{\rlap{\lower4pt\hbox{\hskip1pt$\sim$}}
    \raise1pt\hbox{$>$}}}
\def\fun#1#2{\lower3.6pt\vbox{\baselineskip0pt\lineskip.9pt
  \ialign{$\mathsurround=0pt#1\hfil##\hfil$\crcr#2\crcr\sim\crcr}}}

\input epsf

\def\beq{\begin{equation}}
\def\eeq{\end{equation}}
\def\bea{\begin{eqnarray}}
\def\eea{\end{eqnarray}}

\begin{document}

\preprint{IPMU-10-0080}
\title{Holographic Principle and the Surface of Last Scatter}
\author{Paul Howard Frampton~$^{\bf a,b}$}
\email{paul.frampton@ipmu.jp}
\email{frampton@physics.unc.edu}
\affiliation{$^{\bf a}$ Institute for the Physics and Mathematics of the Universe, University of Tokyo, Kashiwa, 
Chiba 277-8568, Japan} 
\affiliation{$^{\bf b}$ Department of Physics and Astronomy, University of North Carolina,
Chapel Hill, NC 27599-3255, USA.}

\bigskip
\bigskip
\bigskip
\bigskip
\bigskip
\bigskip
\bigskip

\begin{abstract}
Using data, provided by WMAP7, I calculate the entropy
of the visible universe, where visible refers to electromagnetic radiation, and 
hence the visible universe is bounded 
by the Surface of Last Scatter. The dimensionless
entropy, $S/k$, is $(8.85 \pm 0.37)$ times larger than
allowed by a simplified and non-covariant version
of the holographic principle, that the entropy cannot exceed that of a black hole.
The measurement of a shift parameter, 
introduced by Bond, Efstathiou and Tegmark in 1997,
plays an important role in the accuracy of the calculation, which leads to the 
large discrepancy.

\end{abstract}
\maketitle

\section{Introduction}

\bigskip

\noindent
An interesting and profound idea about the degrees of freedom
describing gravity, is the holographic principle
\cite{Hooft,Susskind}. 

\bigskip

\noindent
For the case of a sphere, with mass $M$,  of radius $R$, where $R$ will be the co-moving
radius for the expanding universe, a simplified, and non-covariant form, of
the holographic principle, states
that the entropy, $S/k$, has an upper limit equal to that of a black hole, {\it i.e.}

\bigskip

\begin{equation}
\left( \frac{S}{k} \right) \leq \left( \frac{S}{k} \right)_{BH}
= \left( \frac{ 4\pi R_S^2}{l_{Planck}^2} \right) 
\label{HP}
\end{equation}

\bigskip

\noindent
where $G$ is Newton's
constant, $R_S = 2 G M$ is the Schwarzschild radius and $l_{Planck}$
is the Planck length. It should be emphasized that Eq.(\ref{HP}), to
which counterexamples are rife, is not the same
as the generally covariant holographic principle, enunciated in terms
of null hypersurfaces, in \cite{Bousso}, as a 
generalization of Eq. (\ref{HP}). Nevertheless, it is interesting,
from the viewpoint of the physical understanding of the
visible universe, to use accurate observational data to
check, whether the simplied, and non-covariant, Eq.(\ref{HP})
is satisfied at the present time, $t=t_0$, and in the past,
cognizant that, with dark energy,
if $R$ sufficiently increases, Eq.(\ref{HP})
might, in any case, eventually be violated. Note that, at the time
of \cite{Hooft}, before
dark energy, if Eq.(\ref{HP}) is now satisfied, 
one might expect it to remain so.

\bigskip

\noindent
The inequality, Eq.(\ref{HP}), is believed to be saturated by a black hole,
although there is no experimental evidence, for such a statement.

\bigskip

\noindent
The holographic principle is supported, by string theory. The AdS/CFT correspondence
\cite{AdSCFT} is an explicit realization of Eq.(\ref{HP}), and so, apart from the non-trivial subtlety that our universe is dS, not AdS, from the viewpoint
of string theory, there is every reason to believe the covariant
holographic principle, and to wish to check Eq.(\ref{HP}).
It is related to recent considerations of the entropy of the universe
\cite{V,EFS1,EFS2}.

\bigskip

\noindent
However, physics is an empirical science, and therefore the scientific method
dictates that we should find a physical example, in which Eq.(\ref{HP})
can be calculated. The result, reported here, is that a detailed and accurate check of Eq.(\ref{HP}), as applied to
the visible universe, fails, by a statistically-significant amount, although
in the past, a few billion years ago, it was satisfied.

\bigskip

\noindent
I should define, precisely, what is meant by the visible universe. It is the
sphere, centered for convenience at the Earth, and with a radius
$d_A(Z^*) = 14.0 \pm 0.1 Gpc$. The value of $d_A(Z^*)$ is the particle horizon corresponding
to the recombination red shift $Z^* = 1090 \pm 1$, and is measured directly by WMAP7 \cite{WMAP7},
without needing the details of the expansion history. Thus, "visible" means
with respect to electromagnetic radiation.

\section{The Visible Universe}

\bigskip

\noindent
The motion is that the visible universe, so defined, is a physical object
which should be subject to the holographic principle. It is an expanding,
rather than a static, object, yet my understanding is that the principle,
at least in its covariant form,
is still expected to be valid.

\bigskip

\noindent 
I shall use the notation employed by the WMAP7 paper \cite{WMAP7},
from which all observational data are taken. 

\bigskip

\noindent
The present age, $t_0$, of the universe
is measured to be

\begin{equation}
t_0 = 13.75 \pm 0.13 Gy
\label{age}
\end{equation} 

\bigskip

\noindent
The comoving radius, $d_A(Z^*)$,  of the visible universe, is,
likewise, measured to one percent accuracy, as

\bigskip

\begin{equation}
d_A(Z^*) \equiv (1 + Z^*) D_A(Z^*) = c\int^{t_0}_{t^*} \frac{dt}{a(t)} =14.0 \pm 0.1 Gpc
\label{d}
\end{equation}

\bigskip

\noindent 
where it is noted that the measurement, of $d_A(Z^*)$,
does not require knowledge, of the expansion history, $a(t)$, for $t^* \leq t \leq t_0$.

\bigskip

\noindent
The critical density, $\rho_c$, is provided by the formula

\bigskip

\begin{equation}
\rho_c = \left( \frac{3 H_0^2}{8 \pi G} \right)
\label{critical}
\end{equation}

\bigskip

\noindent
whose value depends  on $H_0$, as does the total, baryonic plus dark,
matter density, $\rho_m$

\bigskip

\begin{equation}
\rho_m \equiv \Omega_m \rho_c
\label{matter}
\end{equation}

\bigskip

\noindent
Because the error on the Hubble parameter, $H_0$,
is several per cent, it is best to avoid $H_0$, in checking the holographic principle.

\bigskip

\noindent
The mass of the matter, M(Z*), contained in the visible universe, is

\bigskip

\begin{equation}
M(Z^*) = \frac{4 \pi}{3} d_A(Z^*)^3 \rho_m
\label{M}
\end{equation}

\bigskip

\noindent
and the Schwarzschild radius, $R_S(Z^*)$, is given by

\bigskip

\begin{equation}
R_S(Z^*) \equiv 2 G M(Z^*)
\label{Schwarzschild}
\end{equation}

\bigskip

\noindent
Collecting results enables the desired accurate check of the 
simplified holographic principle,
which compares entropy, $S/k$, for the visible universe, $(S/k)_{V.U.}$, 
with entropy, $(S/k)_{B.H.}$, for a black hole, of the same mass. According to
Eq.( {\ref{HP}), this requires

\bigskip

\begin{equation}
\left[\frac{\left(\frac{S}{k}\right)_{V.U.}}{\left(\frac{S}{k}\right)_{BH}} \right]  \leq 1
\label{HPuniverse}
\end{equation}

\bigskip

\noindent
A shift parameter, $R$, was defined by Bond, Efstathiou and Tegmark (BET)
in \cite{Bond}, as

\bigskip

\begin{equation}
R = \frac{\sqrt{\Omega_m H_0^2}}{c} (1+Z^*) D_A(Z^*)
\label{shift}
\end{equation}

\bigskip

\noindent
which was, with great prescience, introduced by BET, as a dimensionless
quantity, to be measured, accurately, by CMB observations.

\bigskip

\noindent
This BET shift parameter, $R$, of Eq. (\ref{shift}), is given in \cite{WMAP7},
as

\bigskip

\begin{equation}
R = 1.725 \pm 0.018
\label{R}
\end{equation}

\bigskip

\noindent
A little algebra shows that the BET shift parameter $R$ 
provides the most accurate 
available check, of the holographic principle, by virtue of the result

\bigskip

\begin{equation}
\left[\frac{\left(\frac{S}{k}\right)_{V.U.}}{\left(\frac{S}{k}\right)_{BH}} \right]  \equiv R^4 = 8.85 \pm 0.37
\label{HPcheck}
\end{equation}

\bigskip

\noindent
showing a violation, by $21\sigma$, of Eq.({\ref{HPuniverse}).

\section{Discussion}

\noindent
To my knowledge, the visible universe is, at present, the only physical object,
for which it is possible to calculate, and compare with experiment, or observation,
the simplified holographic principle.

\bigskip

\noindent
From Eq. (\ref{HPuniverse}), the radius $d_A(Z_{HP}) = (1+Z_{HP})D_A(Z_{HP})$,
at which the violation of Eq.(\ref{HP}), begins, is $d_A(Z_{HP}) = 8.4\pm0.1Gpc$,
at a time, comparable to when the cosmic deceleration ends, and becomes acceleration.
This is strongly supportive of the idea of an entropic accelerating universe,
as discussed in \cite{EFS1}.

\bigskip

\noindent
The original aim, of the present work, was to confirm, at $t=t_0$, 
the inequality, Eq.(\ref {HPuniverse}). It was, therefore, surprising to learn that it is violated, with high statistical significance, and has been so, for billions of years.

\newpage

\section*{Acknowledgements}

\noindent
This work was supported by the World Premier International Research Center Initiative 
(WPI initiative) MEXT, Japan, and by U.S. Department
of Energy under Grant No. DE-FG02-05ER41418.

\end{document}